\documentclass{emulateapj}
\usepackage{amsmath}
\usepackage{natbib}

\begin{document}

\bibliographystyle{apj}

\title{{\tt Exo-Transmit}: An Open-Source Code for Calculating Transmission Spectra for Exoplanet Atmospheres of Varied Composition}

\author{Eliza M.-R. Kempton}

\affil{Department of Physics, Grinnell College, Grinnell, IA 50112}

\email{kemptone@grinnell.edu}

\author{Roxana Lupu}

\affil{BAER Institute/NASA Ames Research Center, Moffet Field, CA 94035, USA}

\author{Albert Owusu-Asare}

\affil{Grinnell College, Grinnell, IA 50112}

\author{Patrick Slough}

\affil{Grinnell College, Grinnell, IA 50112}

\author{Bryson Cale}

\affil{Grinnell College, Grinnell, IA 50112}

\begin{abstract}

We present \texttt{Exo-Transmit}, a software package to calculate exoplanet transmission spectra for planets of varied composition.  The code is designed to generate spectra of planets with a wide range of atmospheric composition, temperature, surface gravity, and size, and is therefore applicable to exoplanets ranging in mass and size from hot Jupiters down to rocky super-Earths.  Spectra can be generated with or without clouds or hazes with options to (1) include an optically thick cloud deck at a user-specified atmospheric pressure or (2) to augment the nominal Rayleigh scattering by a user-specified factor.  The \texttt{Exo-Transmit} code is written in C and is extremely easy to use.  Typically the user will only need to edit parameters in a single user input file in order to run the code for a planet of their choosing.  \texttt{Exo-Transmit} is available publicly on Github with open-source licensing at \url{https://github.com/elizakempton/Exo_Transmit}.

\end{abstract}


\section{Introduction \label{intro}}

To date, constraints on exoplanet atmospheric composition have come primarily from transmission spectra -- the wavelength-dependent absorption of starlight that occurs during primary transit for a planet passing in front of its host star.  The transmission spectrum results from a deeper transit occurring at wavelengths where the atomic and molecular species that comprise the exoplanet atmosphere have high opacity.  Robust constraints on the atmospheric composition of an exoplanet were first garnered from transmission spectrum observations of HD 209458b \citep{cha02}. A minutely deeper transit at wavelengths corresponding to the sodium resonance doublet allowed for limits to be placed on the sodium abundance in the planet's atmosphere.  Somewhat surprisingly, the effect was smaller than what had been predicted from modeling of hot Jupiter transmission spectra, leading the authors to conclude that HD 209458b had a lower than expected abundance of sodium gas (either due to low primordial abundances or a chemical depletion effect), or that high obscuring clouds or hazes were present.  

Since those initial pioneering observations, transmission spectroscopy has remained the primary method by which the composition of exoplanet atmospheres has been determined.  Transmission spectra have provided positive identifications of a variety of atoms and molecules in the atmospheres of transiting exoplanets including water \citep{dem13, wak13, kre14, mcc14, kre15, eva16}, carbon monoxide \citep{sne10, bro16}, sodium \citep[e.g][]{cha02, nik14, wyt15}, potassium \citep{sin11, sin15}, and calcium \citep{ast13}.  Furthermore, atmospheric aerosols in the form of clouds or haze have been identified by their effects on exoplanet transmission spectra \citep[e.g.][]{kre14, sin16} -- typically as gray or gently sloped opacity sources that reduce the expected strength of otherwise prominent atomic and molecular spectral features.  

Radiative transfer models of transmission spectra have been key to interpreting the observational results.  Forward models of giant planet atmospheres \citep[e.g.][]{bur97, sea00, bar01, for03} typically start with a one-dimensional (1-D) radiative-convective temperature-pressure (T-P) profile that has been calculated self-consistently with the underlying atmospheric composition using a radiative transfer solver to ensure that energy balance is maintained throughout the atmosphere.  Assuming this uniform temperature structure persists along any radial trajectory through the atmosphere, the transmission spectrum is then generated by determining the fraction of starlight transmitted through the atmosphere along sightlines that intersect an Earth-bound observer.  Care must be taken to correctly account for the path length of these oblique rays to determine the line-of-sight optical depth (often called the ''slant optical depth'').  The fraction of starlight that passes fully through the atmosphere and out the other side, determined on a wavelength-by-wavelength basis, is the transmission spectrum.  Calculations of transmission spectra for planets with non-uniform temperature structure based on 3-D general circulation models (GCMs) typically show good agreement with the results from the aforementioned 1-D models \citep{for10, bur10}, motivating the simpler modeling approach for interpreting data at the level of observational precision that can currently be attained.  More recently, atmospheric retrieval codes have been employed to extract atmospheric parameters and abundances from observations of exoplanet transmission spectra \citep{mad09, ben12, lee14, lin16}.  As opposed to the forward models, which predict the transmission spectrum based on assumptions about the atmospheric composition and equilibrium temperature, retrieval codes back out the composition and T-P profile by generating large suites of models and using statistical techniques to determine the best-fit parameters.  

Models of giant planet atmospheres have generally assumed solar or near-solar composition for the atmosphere, in line with predictions of core accretion theory.  Models of the transmission spectra of low-mass exoplanets have been developed more recently that contend with the greater diversity of atmospheric composition expected of these objects.  The secondary outgassed atmospheres of super-Earths (planets in the 1-10 M$_{\oplus}$ range) are predicted to have highly varied atmospheric composition, with end-member outcomes including predominantly H$_2$O, CO$_2$, or H$_2$ gas \citep{elk08, sch10}.  Consequently, models designed to generate super-Earth transmission spectra \citep{mil09, ben12, how12} are able to incorporate an appropriately broad range of atmospheric composition as input.  Unlike thermal emission spectra, transmission spectra are not strongly sensitive to atmospheric temperature gradients \citep[e.g.][]{mil09}. Therefore, while integrating more diverse atmospheric composition, the aforementioned models of super-Earth transmission spectra typically simplify the computationally intensive portion of the calculation associated with self-consistently generating the radiative-convective T-P profile.  

Here we present a new publicly available open-source code for generating forward models of exoplanet transmission spectra, called \texttt{Exo-Transmit}.  The code is an extension of the super-Earth atmosphere radiative transfer code presented in \citet{mil09} and \citet{mil10} and is designed to generate spectra of planets with a wide range of temperature, surface gravity, size, and atmospheric composition.  While originally intended as a model for low-mass exoplanets,  \texttt{Exo-Transmit} can generate transmission spectra for atmospheres of diverse bulk composition including variants on solar composition, and is therefore suitable for calculating spectra of giant planets as well.  Spectra can be simulated with or without clouds or haze with options to (1) include an optically thick cloud deck at a user-specified atmospheric pressure or (2) to augment the nominal Rayleigh scattering by a user-specified factor.  

\texttt{Exo-Transmit} is written in C and is extremely easy to use.  As an illustration of this point, a sophomore  undergraduate political science major who had never previously encountered a Unix-based operating system recently successfully downloaded and ran the code for an independent study project.  The code is well-documented with a user's manual that is included with the installation.  Typically the user will only be required to edit parameters in a single user input file in order to run the code for a planet of their choosing.   The entire code, including associated opacity and chemistry files, takes up approximately 1.25 GB of hard drive space.  On a single processor, the code takes approximately one minute to run from start to finish for a cloud-free model, with most of that time going to read in the individual opacity files for each atom and molecule.  

In Section~2 we provide a detailed description of \texttt{Exo-Transmit}.  In Section~3 we present results from the model and comment on its validity and applicability across a range of planet types.  Finally, we summarize in Section~4.

\section{Model Description \label{methods}}

\subsection{Overview}

\begin{figure}
\includegraphics[angle=270, scale=0.32]{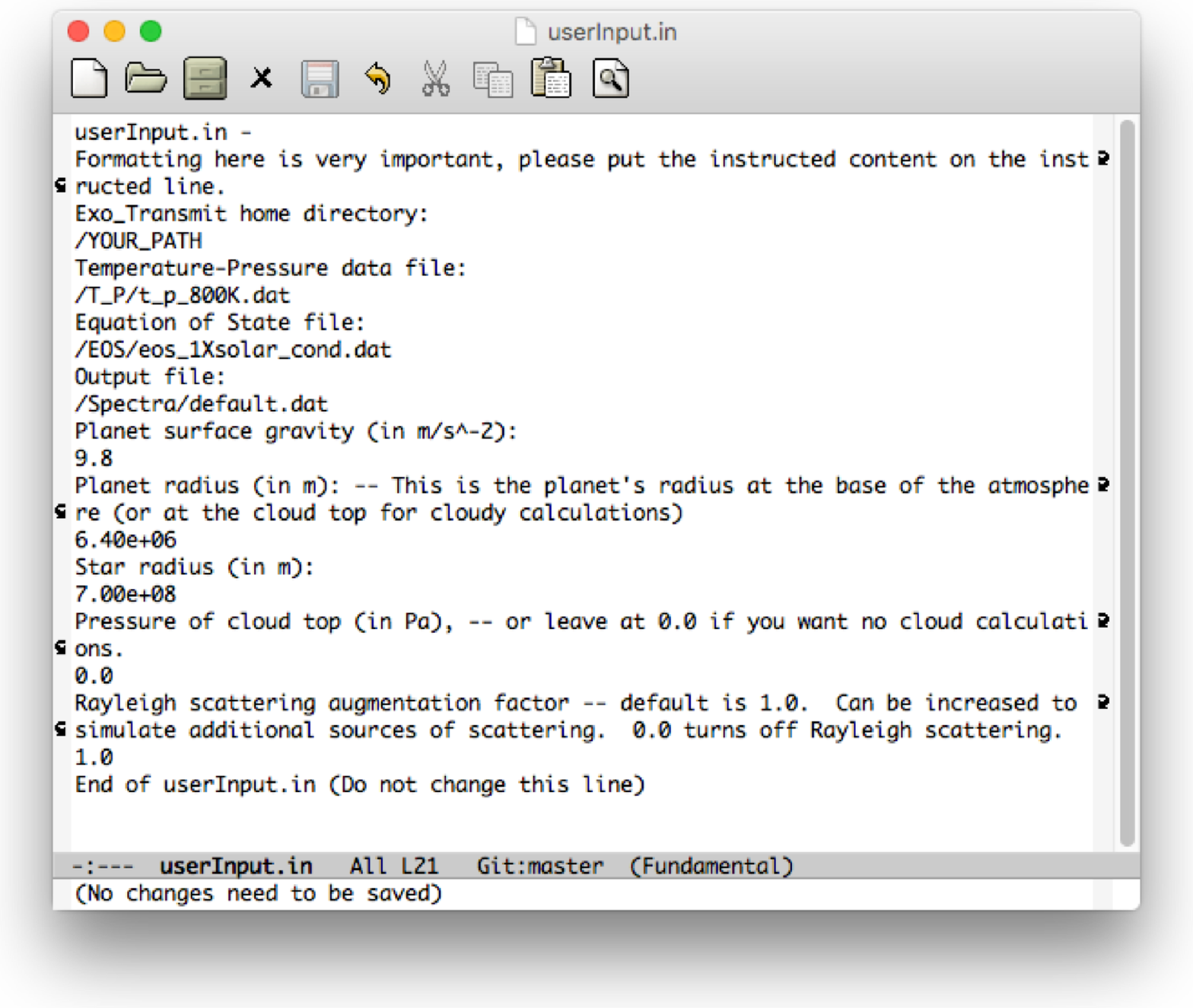}
\caption{User input file for \texttt{Exo-Transmit}.  \label{input_fig}}
\end{figure}

\texttt{Exo-Transmit} is the first publicly available release of the transmission spectrum model described in \citet{mil09} and \citet{mil10}.  The code allows a user to calculate the transmission spectrum of an exoplanet of essentially arbitrary atmospheric composition, where arbitrary here means any combination of the 30 atoms and molecules for which opacity data is included with the code.  For each run of the code, the user provides input information about the planetary and stellar radii, the planet's surface gravity, the 1-D atmospheric T-P profile, the location (in terms of pressure) of any cloud layers, a factor for excess Rayleigh scattering, and the composition of the atmosphere (see Figure~\ref{input_fig}).  The T-P profile and gas phase compositional abundances (provided on a fixed temperature/pressure grid) can be selected from a set of files included with the code, or can be provided externally by the user if the same file format is maintained.  

The code solves the equation of radiative transfer for absorption (only) of starlight passing through the planet's atmosphere  
\begin{equation}
I(\lambda) = I_{0_\lambda} e^{-\tau_{\lambda}}
\end{equation}
where $I$ and $I_{0}$ are the emergent and incident (stellar) intensity, respectively. The line-of-sight optical depth $\tau_{\lambda}$ is calculated according to
\begin{equation}
\tau_{\lambda} = \int\kappa_{\lambda} d\ell
\end{equation}
where $\kappa_{\lambda}$ is the total opacity at wavelength $\lambda$ and $d\ell$ is the differential path length along the observer's line of sight.  While scattering into and out of the beam is not explicitly included in this calculation, Rayleigh scattering is included as an opacity source that contributes to the total $\kappa_{\lambda}$ (see Section~\ref{clouds}).  The wavelength-dependent transit depth, $D_{\lambda}$ is then calculated according to
\begin{equation}
D_{\lambda} = 1 - \frac{F_{in}}{F_{out}} = 1 - \frac{(1 - (\frac{R_{pl, atm}}{R_*})^2)F_{*} + F_{atm}}{F_{*}}
\end{equation}
where $F_{in}$ and $F_{out}$ are the in-transit and out-of-transit flux, respectively, $R_{*}$ is the stellar radius, $R_{pl,atm}$ is the combined radius of the planet plus the entire modeled portion of the atmosphere (which typically extends from 1 bar to a pressure of 1 $\mu$bar in our T-P profiles), and $F_{*}$ is the host star flux.  The additional flux that passes through the planet's atmosphere, $F_{atm}$, is obtained by integrating the intensity in each beam, from Equation~1, over the solid angle subtended by the atmosphere.

A full run of the \texttt{Exo-Transmit} will complete the following set of steps in order:  
\begin{enumerate}
\item The gas phase abundances that set the composition of the atmosphere are read in on a fixed T-P grid.
\item Opacities for the same set of gases are read in individually on a fixed temperature-pressure-wavelength grid.  The T-P component of the grids defined in step (1) and step (2) are required by the code to be identical.   
\item Total opacities are computed from the gas phase mixing ratios on the same T-P grid as step (1).
\item The 1-D T-P profile is read in with the requirement that it must fall entirely within the T-P grid defined in step (1).
\item The radiative transfer calculation for transit geometry is completed according to Equations~1 and 2, and the wavelength-dependent transit depth is calculated according to Equation~3. 
\end{enumerate}
The details of steps 1-4 of this process are laid out in more detail in the following subsections.  

\subsection{Temperature-Pressure Profiles}

\begin{figure}
\includegraphics[angle=90, scale = 0.37]{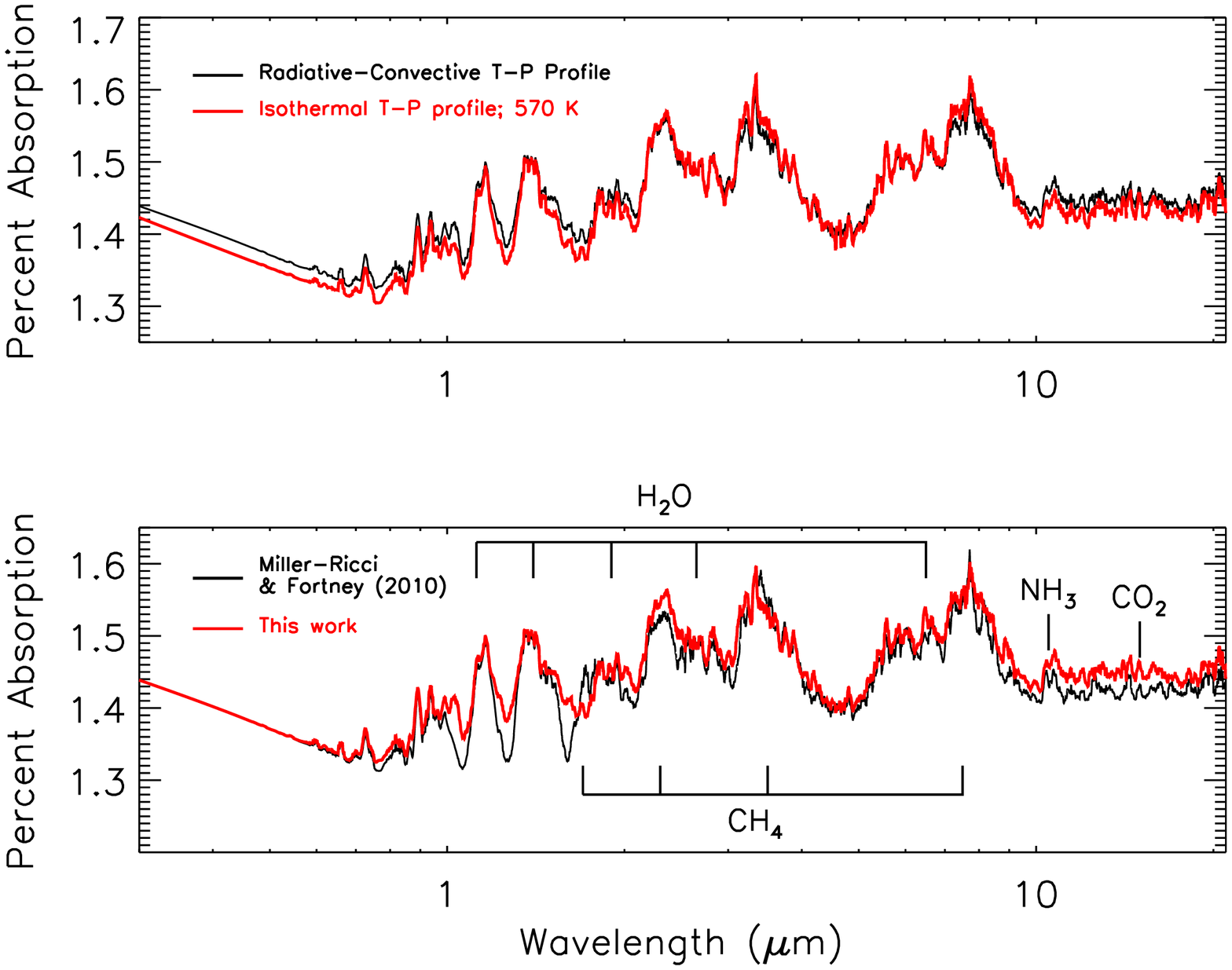}
\caption{Sample transmission spectra for the super-Earth GJ 1214b.  Upper panel: A comparison between solar metallicity spectra generated by \texttt{Exo-Transmit} using the solar composition radiative-convective T-P profile from \citet{mil10} and an isothermal profile with $T = 570$ K.  Lower panel: A comparison between the solar composition GJ 1214b transmission spectrum from \citet{mil10} and an equivalent spectrum generated by \texttt{Exo-Transmit}.  The latter spectrum incorporates updated opacity data, the more recent solar system atomic abundances of \citet{lod03}, and a chemical equilibrium abundance calculation that includes rainout of condensibles.  Of these three effects, the updated opacity data produce the most noticeable changes in the spectrum.   \label{GJ1214b_fig}}
\end{figure}

\texttt{Exo-Transmit} includes a set of isothermal T-P profiles that can be used to generate transmission spectra.  Because they are essentially pure absorption spectra, transmission spectra, unlike thermal emission spectra, are not highly sensitive to vertical temperature gradients in a planet's atmosphere.  They are, however, sensitive to the absolute temperature of the atmosphere in the following two ways.  First, the temperature is a key factor in setting the local scale height, which is in turn primarily responsible for setting the depth of absorption features in the transmission spectrum.  Secondly, the local gas composition and opacity are functions of temperature.  For these reasons, isothermal T-P profiles with the temperature chosen to closely reflect the local temperature at the location where the transmission spectrum originates (typically $\sim$ 1 mbar) are typically sufficient for generating accurate transmission spectra.    As an example, Figure~\ref{GJ1214b_fig} (upper panel) shows the transmission spectrum of GJ 1214b using a T-P profile calculated from a full radiative-convective equilibrium model along with the best-fit spectrum using an isothermal T-P profile, revealing the (minor) level of inconsistency between the two approaches.  

T-P profiles with temperatures between 300 and 1,500 K, with a step size of 100 K are provided with \texttt{Exo-Transmit}.  Spectra for intermediate temperatures can be generated by creating an isothermal T-P profile following the same file format as those included in the code, or can be obtained by interpolating between spectra for bracketing temperatures.  If a more realistic, non-isothermal, T-P profile is preferred, the user can provide one following the same file format as the T-P files included with \texttt{Exo-Transmit}.

\begin{deluxetable*}{lc}
\centering
\tablecaption{Atoms and Molecules Included in \texttt{Exo-Transmit} \label{tab:chemistry}}
\tablewidth{0pt}
\tablehead{
}
\startdata
EOS files & \parbox[c]{4.0in}{C, CH$_4$, CO, COS, CO$_2$, C$_2$H$_2$, C$_2$H$_4$, C$_2$H$_6$, H, HCN, HCl ,HF, H$_2$, H$_2$CO, H$_2$O, H$_2$S, He, K, MgH, N, N$_2$, NO$_2$, NH$_3$, NO, Na, O, O$_2$, O$_3$, OH, PH$_3$, SH, SO$_2$, SiH, SiO, TiO, VO} \\
\\
Molecular Opacities & \parbox[c]{4.0in}{CH$_4$, CO, COS, CO$_2$, C$_2$H$_2$, C$_2$H$_4$, C$_2$H$_6$, HCN, HCl ,HF, H$_2$CO, H$_2$O, H$_2$S, MgH, N$_2$, NO$_2$, NH$_3$, NO, O$_2$, O$_3$, OH, PH$_3$, SH, SO$_2$, SiH, SiO, TiO, VO} \\
\\
Atomic Opacities & Na, K \\
\\
Collision Induced Opacities & \parbox[c]{3.8in}{CH$_4$-CH$_4$, CO$_2$-CO$_2$, H$_2$-H$_2$, H$_2$-He, H$_2$-H, H$_2$-CH$_4$, He-H, N$_2$-CH$_4$, N$_2$-H$_2$, N$_2$-N$_2$, O$_2$-CO$_2$, O$_2$-N$_2$, O$_2$-O} \\
\\

\enddata
\end{deluxetable*}

\subsection{Chemistry}

Equation of state (EOS) files (i.e. abundances for key absorbers and major atmospheric constituents as a function of temperature and pressure) are provided with \texttt{Exo-Transmit} for a range of plausible atmospheric compositions.  Specifically, EOS files have been calculated with metallicity values of 0.1, 1, 5, 10, 30, 50, 100, and 1000 times the solar value, based on the solar system abundances of \citet{lod03}.  We have also included a set of models at solar composition but with varying C-to-O ratios ranging from 0.2 to 1.2 (with 0.5 being the solar system value).  Finally, for each of the molecules for which we have opacity data, we have included an EOS file corresponding to an atmosphere composed wholly of that molecule.  The list of molecules included in each EOS file is provided in Table~\ref{tab:chemistry}, and the corresponding opacity data for this list of molecules is described in the following sub-section.

For the atmospheres of mixed composition, we have used a Gibbs free energy minimization code to determine chemical abundances.  This code, not included with \texttt{Exo-Transmit} but described in detail in \citet{mba16}, calculates abundances in chemical equilibrium of over 550 gas phase and condensed (liquid and solid) species, of which 36 of the gas phase species are reported in our EOS tables (see Table~1).  The rest of the species typically have either very low abundances or negligible opacity in the optical and IR.  We do not include abundance or opacity data for condensed species with \texttt{Exo-Transmit}.  However, to further assess the effects of condensation chemistry on transmission spectra, all of the mixed composition EOS files are provided in two versions --- one in which the Gibbs free energy minimization code has been run to only include gas phase chemistry, disregarding condensation processes, and a second that includes condensation chemistry with rainout, as described in \citet{mba16}.  For the high metallicity and carbon-rich atmospheres, in which condensation of carbon in the form of graphite is expected, a third EOS file is included that accounts for condensation of all species except for graphite.  In our chemical equilibrium calculations with rainout, graphite condensation fully removes carbon at temperatures lower than the condensation temperature, which fully depletes the atmosphere of all carbon-bearing species at low temperature.  Additional work on cloud microphysics is required to assess how physically plausible this scenario is, so we leave the user with options to generate spectra with and without graphite condensation in the meantime.  

All EOS files provided with \texttt{Exo-Transmit} are computed on the same T-P grid of 100 - 3,000 K (in steps of 100 K) and 10$^3$ - 10$^{-9}$ bar (in logarithmic steps of one dex).  We caution that the EOS files have limited use at the very low-temperature and low pressure end of this range because condensation of volatiles into ices and molecular diffusion are not included in our calculations.   As with the T-P profiles, the user of \texttt{Exo-Transmit} may provide their own EOS files, provided they follow the same format and T-P grid as those supplied with the code.  

\subsection{Opacities}

Opacities for 30 individual molecular and atomic species have been tabulated on a fixed temperature-pressure-wavelength grid.  The T-P portion of the grid corresponds exactly with the T-P grid selected for the EOS files.  The wavelength grid extends from 0.1 to 170 $\mu$m at a fixed spectral resolution of 10$^3$.   The line lists used to generate the molecular opacities are the same ones used in \citet[][their Table 2]{lup14} with the following exception --- we do not include CaH, ClO, CrH, FeH, or LiCl because our chemical equilibrium calculations do not currently provide the abundances of these 5 molecules.  Our molecular database \citep{Freedman:2008,Freedman:2014} has been validated by numerous exoplanet and brown dwarf studies and draws from the best available public data, communication with other workers in the field, and our own calculations. The opacities are adequate for temperatures up to a few thousand Kelvin, as checked against the latest laboratory experiments. Spectral lines are broadened with Voigt profiles, assuming H$_2$ broadening. This is appropriate for current exoplanet transit observations, since the gas giants have H$_2$-He dominated atmospheres, and pressure broadening is unlikely to have a significant effect for smaller planets. Atomic opacities are included for Na and K. All atomic and molecular opacities will be updated in the GitHub repository as better data becomes available.  Collision-induced opacities resulting from inelastic collisions of pairs of molecules are calculated for major species to the extent they are available in the literature. Following \citet{lup14}, we are currently using the set released by the HITRAN team \citep{Richard:2012}. All of the opacity sources -- atomic, molecular, and collision-induced-- incorporated into \texttt{Exo-Transmit} are reported in Table~1.  Rayleigh scattering cross sections are calculated separately for each species and then weighted by the individual number densities and summed to produce the total Rayleigh scattering opacity self consistently with the atmospheric composition.  

Total opacities for each temperature-pressure-wavelength grid point are determined by summing all of the individual opacity sources -- the molecular and atomic opacities each weighted by their abundance, the collision-induced opacities weighted by the product of the abundances of the pair of molecules, and the total Rayleigh scattering opacity.  

\subsection{Clouds \label{clouds}}

Observations of exoplanet atmospheres lead us to believe that atmospheric aerosols (clouds and hazes) are commonplace \citep[e.g.][]{sin16}.  Furthermore, all solar system planets and moons with thick atmospheres harbor multiple forms of aerosols\footnote{http://www.planetary.org/connect/our-experts/profiles/sarah-horst.html}.  It is therefore necessary to provide a method for calculating transmission spectra for cloudy and hazy atmospheres, even if the composition and nature of the aerosol is unknown.  \texttt{Exo-Transmit} allows the user to incorporate aerosols into the transmission spectrum calculation following one of two ad-hoc procedures.  The first is to insert a fully optically thick gray cloud deck at a user-specified pressure.  The second is to increase the nominal Rayleigh scattering by a user-specified factor.  

\begin{figure}
\includegraphics[angle=90, scale = 0.37]{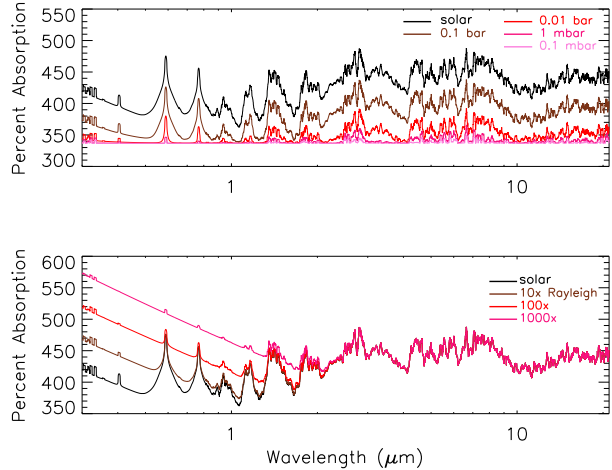}
\caption{Spectra for atmospheres with clouds and hazes.  The top panel shows transmission spectra for atmospheres with a fully gray optically thick cloud deck at pressures between 0.1 bar and 0.1 mbar, as indicated.  The bottom panel shows transmission spectra for atmospheres with excess Rayleigh scattering -- 10 - 1,000 times the solar value, as indicated.  In all cases, the aerosol effects have been added to a baseline 1,000 K isothermal atmosphere model calculated at solar composition for a 2 $R_{\oplus}$ planet orbiting a 1 $R_{\odot}$ star with an Earth-like surface gravity of 10 m/s$^2$.    \label{tr_aerosol_fig}}
\end{figure}

For the gray cloud calculation, the user selects a pressure at which the cloud-top will become optically thick.  This value, specified in SI units, is entered into the user input file (see Figure~1).  When a non-zero cloud pressure is specified within the pressure range of the T-P profile, \texttt{Exo-Transmit} stops reading in the T-P profile for pressures in excess of the specified pressure threshold, and the radiative transfer calculation is only performed for pressures below that of the cloud deck. In the final calculation, this results in transmitted starlight passing through only the portion of the atmosphere above the cloud layer.  Examples of cloudy spectra are shown in Figure~\ref{tr_aerosol_fig}.  The effect of clouds is to reduce the strength of absorption features because light is being transmitted through a smaller portion of the upper atmosphere.

A second option is included to arbitrarily increase the nominal degree of Rayleigh scattering by increasing the scattering cross section by a multiplicative factor specified in the user input file.  Excess scattering (beyond what is expected for an aerosol-free atmosphere) has been reported in observations of exoplanet atmospheres, most notably for HD 189733b \citep[e.g.][]{pon13}.  Its effect is to reduce the strength of atomic spectral features and to alter the slope of the continuum in the optical to near-IR portion of the transmission spectrum.  The increased scattering signal has been attributed to high-altitude and small particle aerosols but might alternatively be explained by stellar activity in the exoplanet host \citep{mcc14}.   \citet{sin16} have shown that the current population of hot Jupiters with transmission spectrum observations is consistent with a continuum of hazy atmospheres.  The spectra imply excess Rayleigh scattering anywhere between the solar value and several thousand times augmented relative to solar, depending on the individual object.  The physical mechanism that generates the haze is still being explored.  With \texttt{Exo-Transmit}, the user can augment the Rayleigh scattering beyond the nominal gas phase amount by a factor of their choice.  Because the Rayleigh scattering cross sections are computed self-consistently for each atmosphere, this means that an augmentation factor of 3 for a 100x solar metallicity atmosphere will result in a level of Rayleigh scattering that is 300 times the solar value.  

\section{Transmission Spectrum Calculations}

\begin{figure*}
\includegraphics[scale = 0.43]{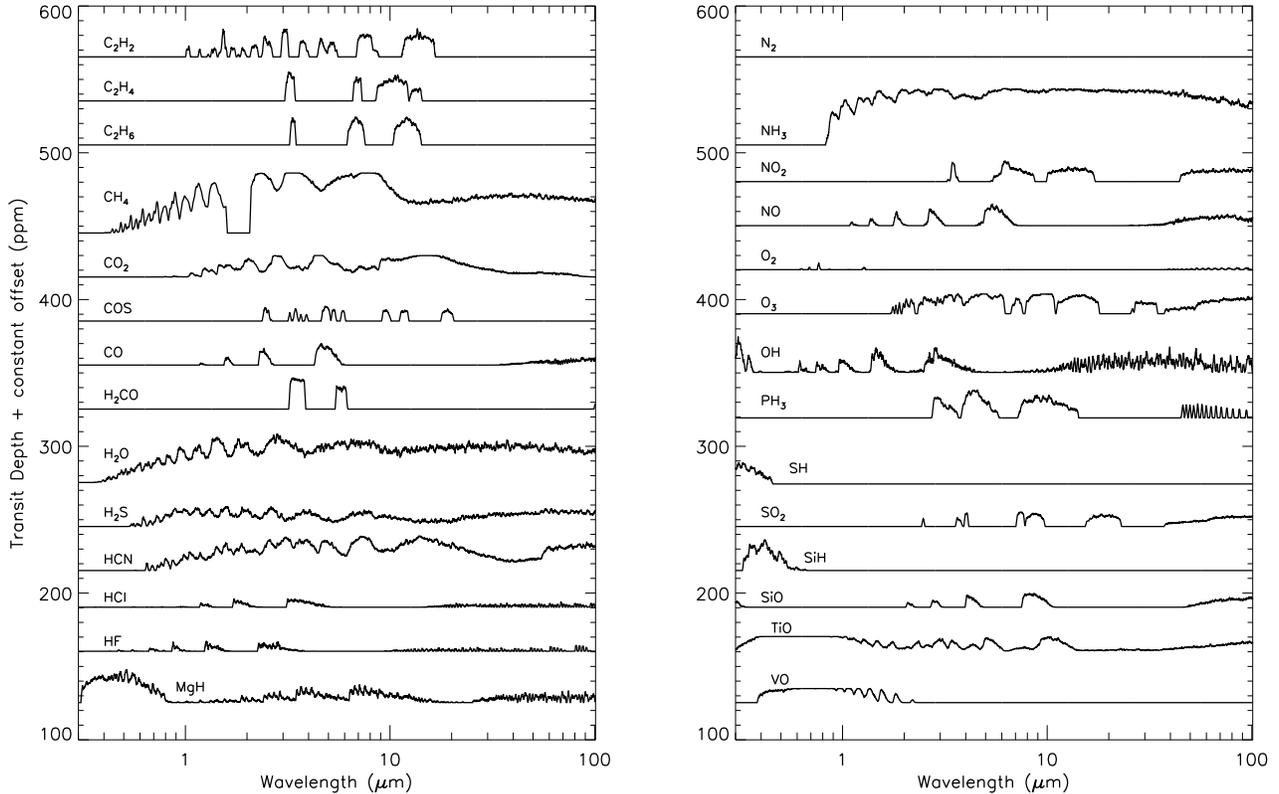}
\caption{Spectra for atmospheres made up entirely of individual molecules.  Each of the molecules with opacity data provided with \texttt{Exo-Transmit} is included on this plot.  The spectra are calculated for a 2 $R_{\oplus}$ planet orbiting a 1 $R_{\odot}$ star with an Earth-like surface gravity of 10 m/s$^2$.  The atmosphere has an isothermal temperature of 1,000 K.    \label{tr_mol_fig}}
\end{figure*}

\begin{figure}
\includegraphics[scale = 0.42]{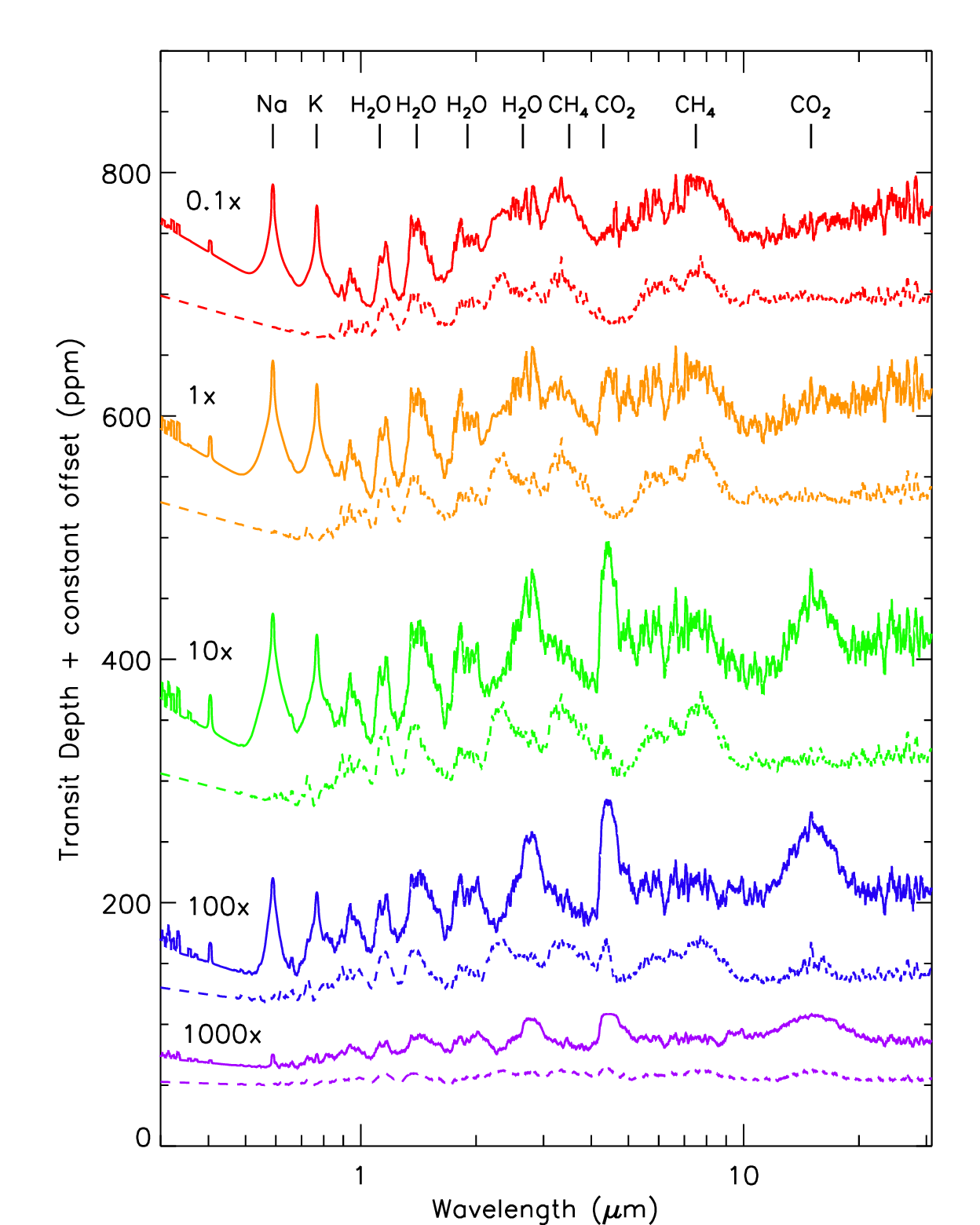}
\caption{Spectra for atmospheres of scaled solar composition.  Metallicities (relative to solar) range from 0.1 times solar (top, red) to 1,000 time solar (bottom, purple), as indicated.  The spectra are calculated for a 2 $R_{\oplus}$ planet orbiting a 1 $R_{\odot}$ star with an Earth-like surface gravity of 10 m/s$^2$.  Solid and dashed lines indicate isothemal atmospheres with $T =$ 1,000 K and  $T =$ 500 K, respectively.    \label{tr_mixed_fig}}
\end{figure}

\begin{figure}
\includegraphics[scale = 0.42]{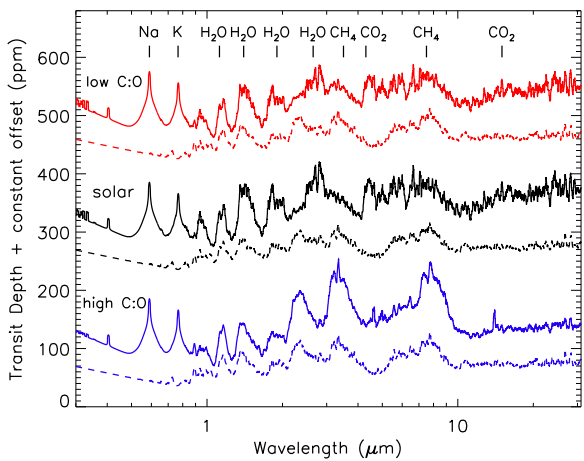}
\caption{Spectra for atmospheres with varying C-to-O ratios -- C:O = 0.2 (top, red), solar C:O (middle, black), and C:O = 1.2 (bottom, blue) , as indicated.  The spectra are calculated for a 2 $R_{\oplus}$ planet orbiting a 1 $R_{\odot}$ star with an Earth-like surface gravity of 10 m/s$^2$.  Solid and dashed lines indicate isothemal atmospheres with $T =$ 1,000 K and  $T =$ 500 K, respectively.    \label{tr_CtoO_fig}}
\end{figure}

Examples of transmission spectra generated by \texttt{Exo-Transmit} are shown in Figures~\ref{tr_aerosol_fig} - \ref{tr_CtoO_fig}.  Figure~\ref{tr_mol_fig} shows the transmission spectra that would be obtained for atmospheres composed fully of each of the 28 molecules, for which opacity data is included with \texttt{Exo-Transmit}.  The scale height is calculated self-consistently for each case, meaning that the absolute strength of spectral features results from a combination of the magnitude of the opacity and the molecular weight of the molecule in question.  For example, TiO and VO are very strong absorbers, but each with a high molecular weight.   Figures~\ref{tr_aerosol_fig}, \ref{tr_mixed_fig},  and \ref{tr_CtoO_fig} show spectra for atmospheres of mixed composition -- solar composition with aerosols, scaled solar composition, and solar composition with varying C:O ratio, respectively.  In all of these figures, condensation chemistry with rainout of condensibles has been included, as evidenced by the lack of sodium and potassium in the 500 K spectra.  We have not allowed for rainout of graphite in any of these spectra, which could notably affect the high metallicity and high C:O atmospheres.  If condensation and complete rainout of graphite were to occur, the transmission spectrum would appear to be completely depleted of carbon-bearing species, and the strong CH$_4$ and CO$_2$ bands would be entirely missing from the spectrum.  This intriguing possibility requires further detailed modeling and may lead to a maximum C:O ratio that is possible to observationally distinguish from exoplanet spectra. 

In Section 2 we have described our choices of opacity and solar abundance data for this work.  We have gone to considerable effort to ensure usage of reliable and up-to-date atomic and molecular information.  However, updates and revisions are consistently being made in the literature to chemical abundance and opacity databases, so it is important to note that the imperfect nature of these data is a key source of uncertainty in models of exoplanet spectra.  To gain an understanding of the effect of different opacity and abundance tables on our modeled spectra, in Figure~\ref{GJ1214b_fig} (lower panel) we show a model of GJ 1214b from \citet{mil10} overlaid with an equivalent model from our current work.  Both spectra were calculated assuming solar abundances and employed an identical set of planetary and stellar parameters.  The \citet{mil10} spectrum used the solar abundances of \citet{asp05}, whereas our current work uses the \citet{lod03} solar system abundances.  Furthermore, the \citet{mil10} spectrum incorporated a more limited set of opacity sources -- H$_2$O, CH$_4$, CO, CO$_2$, and NH$_3$ along with H$_2$ collision-induced opacities -- mostly derived from older opacity databases than the ones employed by \texttt{Exo-Transmit}.  While we believe our current use of updated opacity and abundance tables represents an improvement over previous work, Figure~\ref{GJ1214b_fig} shows the level of discrepancy that one might expect from using different chemical abundance and/or opacity data.  

The absolute transit depth $D$ is given by
\begin{equation}
D = \frac{R_{p}^{2}}{R_{\star}^{2}}.
\end{equation}
where $R_{p}$ and $R_{\star}$ are the planetary and stellar radii.  In cloud-free calculations with \texttt{Exo-Transmit}, $R_{p}$ is the radius of the planet at the base of the atmosphere, which corresponds to a pressure of 1 bar when using the T-P profiles provided with the code.  For calculations with clouds, $R_{p}$ is the radius of the planet at the cloud deck.  The atmosphere will add to the perceived size of the planet, so in most cases, to match the observed transit radius of a planet, the user-selected value of $R_{p}$ must be somewhat smaller than the transit radius.  Alternatively, for small perturbations on the radius, the transit radius can be selected for $R_{p}$, and the entire outputted transmission spectrum can be rescaled by a multiplicative factor close to but less than unity to reproduce the observed transit radius.

The relative depth of transmission spectral features $\Delta D$ for a specified exoplanet are given by
\begin{equation}
\Delta D \sim H = \frac{k T}{\mu g}.
\end{equation}
where $H$ is the pressure scale height, $T$ is the atmospheric temperature, $\mu$ is the mean molecular weight, $g$ is the surface gravity, and $k$ is Boltzmann's constant.  The exact number of scale heights probed by a specific spectral feature depends on both the magnitude of the opacity and on the resolution of the spectrum but is typically between 1 and 10.  

If one wishes to rescale models from \texttt{Exo-Transmit} with differing input parameters, one must perform appropriate mathematical transformations of the relevant parameters, namely $R_{p}$, $R_{\star}$, $T$, and $g$ to reproduce the correct $D$ and $\Delta D$.  The mean molecular weight is calculated internally to \texttt{Exo-Transmit} to be consistent with the local atmospheric composition. For atmospheres of mixed composition (e.g.~solar composition), scaling with temperature should be done cautiously because the chemical abundances can scale non-linearly with temperature.  Especially when crossing the temperature threshold of condensation for a major opacity source, the transmission spectrum can change dramatically across a very small temperature range due to the removal of a key condensible.  While in principle, spectra can also be re-scaled to represent models with slightly different atmospheric composition, we recommend that a new model be calculated from scratch every time a new atmospheric composition is desired.  Both opacities and chemistry can have dramatic and non-linear changes, even when the underlying atmospheric composition is only altered slightly.  

\begin{figure}
\includegraphics[scale=0.48]{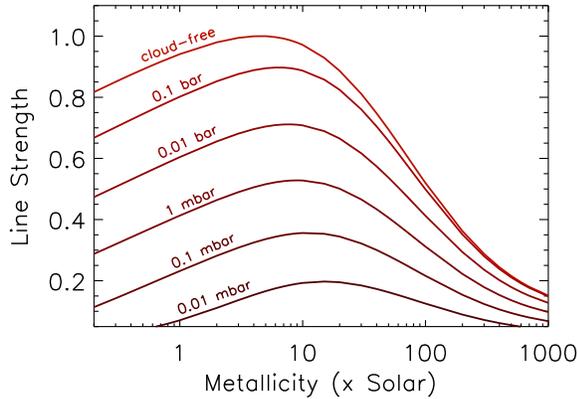}
\caption{Depth of transmission spectral features as a function of metallicity, scaled relative to solar.  The y-axis has been normalized to produce a maximum value of unity.   Spectra have been generated for cloud-free atmospheres as well as atmospheres with clouds at pressures between 0.1 bar and 0.01 mbar, as indicated.  Line strengths were calculated by taking the difference between the maximum and minimum transit depth over the wavelength range $\lambda =  0.5 - 2 $ $\mu$m.  Low metal abundance, high mean molecular weight, and the presence of clouds all reduce the strength of spectral features.  The line strength is degenerate with both metallicity and the cloud deck location.  \label{metal_fig}}
\end{figure}

To demonstrate the various effects on $\Delta D$, Figure~\ref{metal_fig} shows the relative depth of transmission spectrum features obtained for models of varying metallicity and cloud deck location.   We find that the depth of transmission features reaches its peak value at approximately 5 times the solar metallicity for cloud-free models, which sets the maximum possible strength of spectral features in transmission.  At lower metallicity, the abundance of key absorbers falls off dramatically, and with it their associated opacities.  At higher metallicity, the increase in mean molecular weight dominates, causing the depth of spectral features to diminish.  Models with clouds follow the same trend but with smaller $\Delta D$ for increasingly higher cloud layers.  While a number of authors \citep[e.g.][]{mil09, ben12, dew13} have argued for the depth of transmission spectral features being useful as a proxy for measuring key properties of an exoplanet and its atmosphere -- mainly composition and surface gravity -- Figure~\ref{metal_fig} shows the significant degeneracies that exist in interpreting $\Delta D$ to a specified atmospheric composition.  

\section{Summary}

We have developed \texttt{Exo-Transmit}, a software package to generate transmission spectra for exoplanet atmospheres of varied composition, and have made it openly available.  The code can calculate transmission spectra for exoplanets of arbitrary size and surface gravity for a wide range of atmospheric composition.  The novel aspects of \texttt{Exo-Transmit} are (1) its flexibility to generate transmission spectra over a broad range of parameter space associated with known transiting exoplanets ranging from hot Jupiters down to terrestrial exoplanets and (2) its ease of use.  The code can be run as-is using the T-P profiles and EOS files provided with the installation, or the user can alternatively provide their own files following the same format.  \texttt{Exo-Transmit} is written in C and is available via Github at the following link: \url{https://github.com/elizakempton/Exo_Transmit}.  The code is designated as open source under the GNU Free Documentation License.

\acknowledgements
We thank Rostom Mbarek, Laura Schaefer, Thomas Katucki, Zach Berta, Hannah Diamond-Lowe and Natasha Batalha for testing and providing feedback on \texttt{Exo-Transmit} in advance of its public release.  We thank the members of the Grinnell College Computer Science Department, specifically Sam Rebelsky, Henry Walker, and John Stone, for providing Linux and programming support that made this work possible.  We thank the Grinnell College Mentored Advanced Project (MAP) program for supporting the work of A.O.-A., P.S., and B.C.  The work of E.K. was supported by the NASA Planetary Atmospheres program, grant NNX14AP90A and the Grinnell College Harris Faculty fellowship. RL acknowledges support of the NASA Exoplanet Research Program.

\bibliography{ms}

\end{document}